\documentclass[twocolumn,aps,superscriptaddress]{revtex4}
\usepackage{amssymb}
\usepackage{amsfonts}
\usepackage{amsmath}
\usepackage{graphicx}

\begin{document}

\title{Magnetic ratchet effect in bilayer graphene}

\author{Narjes Kheirabadi}
\affiliation{Physics Department, Lancaster University, Lancaster, LA1 4YB, UK}

\author{Edward McCann}
\affiliation{Physics Department, Lancaster University, Lancaster, LA1 4YB, UK}

\author{Vladimir~I.~Fal’ko}
\affiliation{National Graphene Institute, The University of Manchester, Manchester, M13 9PL, UK}

\begin{abstract}
We consider the orbital effect of an in-plane magnetic field on electrons in bilayer
graphene, deriving linear-in-field contributions to the low-energy Hamiltonian arising from
the presence of either skew interlayer coupling or interlayer potential asymmetry, the latter being tunable by an external metallic gate. To illustrate the relevance of such terms,
we consider the ratchet effect in which a dc current results from the application of an
alternating electric field in the presence of an in-plane magnetic field and inversion-symmetry breaking.
By comparison with recent experimental observations in monolayer graphene
[C. Drexler {\em et al}, Nature Nanotech. {\bf 8}, 104 (2013)],
we estimate that the effect in bilayer graphene can be two orders of magnitude greater
than that in monolayer, illustrating that the bilayer is an ideal material for the realization
of optoelectronic effects that rely on inversion-symmetry breaking.
\end{abstract}

\maketitle

\section{Introduction}

\begin{figure}[ht]
   \centering
   \includegraphics[scale=0.6]{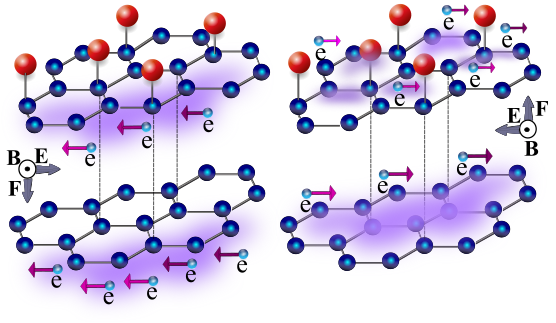}
   \caption{Schematic of bilayer graphene (blue circles) illustrating the ratchet effect in the presence of an in-plane magnetic field $\mathbf{B}$ (out of the page), an alternating electric field $\mathbf{E}$ (to the right or the left) and layer asymmetry illustrated by impurities (red circles) on the upper layer. Electrons are driven towards the lower or upper layer by the Lorentz force, resulting in a relatively high or low mobility, depending on the presence of impurities.
   }
    \label{fig:ratchet}
\end{figure}

The magnetic ratchet effect is a generic feature of inversion-asymmetric two-dimensional systems~\cite{falko89,tara08,tara11,nalitov12,drex13,budkin14,gan14,olbr16}.
It consists of the production of a dc electric current in response to
a steady in-plane magnetic field $\mathbf{B}$ and an alternating electric field $\mathbf{E}$, in the presence of inversion asymmetry. It is illustrated in Fig.~\ref{fig:ratchet} for bilayer graphene where the asymmetry is caused by a larger density of impurities on the upper layer.
For a given direction of electric field (to the right as shown in the left side of Fig.~\ref{fig:ratchet}), electrons are driven downwards by the Lorentz force towards the lower layer where, owing to an absence of impurities, the mobility is relatively high. When the electric field alternates to the opposite direction, electrons are driven upwards towards the upper layer where, owing to the presence of impurities, the mobility is relatively low.
Asymmetry in mobility depending on the direction of electron motion leads to the presence of a non-zero dc current \cite{tara11,drex13}.
Recently, experimental observation of the magnetic ratchet effect has been reported in monolayer graphene
with symmetry broken by the presence of adatoms \cite{drex13} or a superlattice \cite{olbr16}
(for a review of nonlinear optical and optoelectronic effects in graphene see Ref.~\cite{glazgan}).

Here, we contend that bilayer graphene is a natural system in which inversion symmetry may be broken and, thus, in which to observe the magnetic ratchet effect.
Bilayer graphene displays fascinating electronic properties including the presence of chiral quasiparticles \cite{novo06,mccfal06,mcckos13} and the possibility to tune a gap between the conduction and valence bands using potential asymmetry of the layers~\cite{mccfal06}.
Thus, we are able to describe two different mechanisms to break symmetry leading to the ratchet effect:
either a different density of impurities on the two layers of the bilayer or interlayer-symmetry breaking due to the presence of an external gate. By comparison with the analysis of Ref.~\cite{drex13}, we predict the ratchet effect to be up to two orders of magnitude greater in bilayer than in monolayer graphene.

In the presence of an alternating electric field with components $E_x$, $E_y$,
and a steady in-plane magnetic field with components $B_x$, $B_y$, the dc current may
be expressed \cite{tara11} as
\begin{eqnarray}
J_x &=& M_1 \!\!\left[ B_y ( \left| E_x \right|^2-\left| E_y \right|^2)
-B_x (E_xE_y^*+E_yE_x^*) \right] \nonumber \\
&& \qquad + M_2 B_y\left| E \right|^2 + i M_3 B_x (E_xE_y^*-E_yE_x^*), \nonumber \\
J_y &=& M_1 \!\!\left[ B_x(\left| E_x \right|^2-\left| E_y \right|^2) + B_y(E_xE_y^*+E_yE_x^*)\right] \nonumber\\
&& \qquad - M_2B_x\left| E \right|^2 + i M_3 B_y (E_xE_y^*-E_yE_x^*) .\label{jden}
\end{eqnarray}
Here, the functions $M_1$, $M_2$, $M_3$ describe the current response
to different polarizations of light: $M_2$ characterizes the effect of unpolarized light,
$M_1$ includes additional terms that appear if the light is linearly polarized,
$M_3$ describes additional terms that occur for circular polarization with the sign of the current parallel to the magnetic
field dependent on the sense of rotation of the electric field \cite{drex13}.

\section{Methodology}

\subsection{Bilayer graphene in the presence of an in-plane magnetic field}

\subsubsection{Full four-component bilayer model}

\begin{figure}[ht]
   \centering
   \includegraphics[scale=0.35]{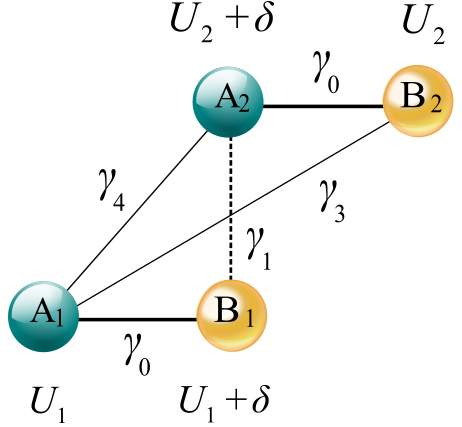}
   \caption {
   Schematic of the unit cell of bilayer graphene with atoms A1, B1 on the lower layer,
   A2, B2 on the upper layer. Straight lines indicate intralayer coupling $\gamma_0$,
   vertical interlayer coupling $\gamma_1$ and skew interlayer couplings $\gamma_3$, $\gamma_4$. Parameters $U_1$, $U_2$, $\delta$ indicate different on-site energies, as described in the main text.}
    \label{fig:unit-cell}
\end{figure}

We consider bilayer graphene with atomic sites A1, B1 on the lower layer,
A2, B2 on the upper layer, and we take into account one $p_z$ orbital per site.
Sites $B1$ and $A2$ lie directly above or below each other, and their orbitals are relatively-strongly coupled, parameterized by interlayer coupling $\gamma_1$. As a result,
$B1$ and $A2$ are referred to as `dimer' sites.
The tight-binding model of bilayer graphene has been studied previously \cite{mccfal06,gui06,nil08}
(for a review see Ref.~\cite{mcckos13}), here we add the effect of an in-plane magnetic field, $\mathbf{B} = (B_x,B_y,0)$, with corresponding
vector potential $\mathbf{A} = z ( B_y , -B_x , 0)$,
chosen to preserve translation symmetry in the graphene plane.
Here $z$ is the Cartesian coordinate perpendicular to the graphene, the lower layer is located
at $z = -d/2$, the upper layer at $z= d/2$, $d$ is the interlayer spacing.
The in-plane field enters as a phase given by a path integral of the
vector potential. For example, the matrix element describing in-plane hopping between an $A$ atom
and three nearest-neighbour $B$ atoms is given by
\begin{eqnarray*}
H_{AB} = - \gamma_0 \sum_{j = 1}^3 \exp \! \left(\frac{i \mathbf{p}}{\hbar}.\left( \mathbf{R}_{Bj} - \mathbf{R}_{A}\right) - \frac{i e }{\hbar}\int_{\mathbf{R}_{Bj}}^{\mathbf{R}_{A}}
\mathbf{A} . \mathbf{d \ell} \right)  ,
\end{eqnarray*}
where $\gamma_0$ is a tight-binding parameter.
Taking such a modification of the tight-binding matrix elements into account,
the electronic Hamiltonian in the vicinity of the Brillouin zone corners (the K points)
may be written, in a basis of A1, B1, A2, B2 sites, as
\begin{eqnarray}
H = \left(
      \begin{array}{cccc}
        U_1 & v\pi_1^{\dagger} & - v_4\pi^{\dagger} & v_3\pi \\
        v\pi_1 & U_1+\delta & \gamma_1 & - v_4\pi^{\dagger} \\
        - v_4\pi & \gamma_1 & U_2+\delta & v\pi_2^{\dagger} \\
        v_3\pi^{\dagger} & - v_4\pi & v\pi_2 & U_2 \\
      \end{array}
    \right) \, . \label{h4}
\end{eqnarray}
Here $v = \sqrt{3} a \gamma_0 / (2\hbar)$ characterises the strength of in-plane
nearest-neighbour A1-B1, A2-B2 hopping, $a$ is the lattice constant,
$\gamma_1$ describes vertical interlayer coupling,
$v_3 = \sqrt{3} a \gamma_3 / (2\hbar)$ characterizes the strength of skew interlayer
A1-B2 hopping, and
$v_4 = \sqrt{3} a \gamma_4 / (2\hbar)$ characterizes the strength of skew interlayer
A1-A2, B1-B2 hopping.
Parameters $U_1$, $U_2$ are the on-site energies of the two layers and $\delta$ describes an energy difference between sites which have neighboring atoms directly above or below them (dimer sites) and those sites which do not \cite{dressel02,nil08,zhang08,li09,mcckos13}.
For in-plane momentum $\mathbf{p} =  ( p_x , p_y , 0)$, the complex momentum operators are $\pi_1$ for the lower layer,
$\pi_2$ for the upper layer and $\pi$ for skew interlayer hopping:
\begin{eqnarray*}
\pi &=& \xi p_x + i p_y \, , \\
\pi_1 &=& \xi ( p_x - b_y ) + i ( p_y + b_x ) \, , \\
\pi_2 &=& \xi ( p_x + b_y ) + i ( p_y - b_x ) \, ,
\end{eqnarray*}
where $b_x = e d B_x /2$, $b_y = e d B_y / 2$, and $\xi = \pm 1$ is an index for the two non-equivalent valleys at wave vectors $\xi (4\pi/3a, 0)$.

\subsubsection{Two-component reduced low-energy Hamiltonian}

As there are four orbitals in the unit cell, Hamiltonian (\ref{h4}) describes four electronic
bands. Due to the relatively-strong interlayer coupling $\gamma_1$ between the
$B1$ and $A2$ dimer sites, the bands associated with their orbitals are split away from
zero energy by $\pm \gamma_1$, while two bands related to the orbitals on the $A1$ and $B2$ sites
touch at zero energy and are approximately quadratic with energy $\epsilon = v^2 p^2/\gamma_1$
\cite{mccfal06,mcckos13}. In order to describe electronic behavior at low energy (lower than $|\gamma_1|$), it is possible to consider a two-component Hamiltonian based on the orbitals on sites $A1$ and $B2$, obtained by eliminating the components related to the dimer sites $B1$ and $A2$. To do this, we follow the procedure detailed previously \cite{mccfal06,mcckos13} taking into account the presence of the in-plane magnetic field. Thus, in a basis of A1, B2 sites, we find:
\begin{eqnarray}
H &=& - \frac{v^2}{\gamma_1} \!\!\left[ 1 + \left(\frac{v_4}{v} + \frac{\delta}{\gamma_1}\right)^2 - \frac{(U_1^2+U_2^2)}{2\gamma_1^2} \right]\!\!\left(
                             \begin{array}{cc}
                               0 & \left(\pi^{\dagger}\right)^2 \\
                               \pi^2 & 0 \\
                             \end{array}
                           \right) \nonumber \\
&& \, + \left(
       \begin{array}{cc}
         U_1 & 0 \\
         0 & U_2 \\
       \end{array}
     \right) + v_3 \left(
       \begin{array}{cc}
         0 & \pi \\
         \pi^{\dagger} & 0 \\
       \end{array}
     \right) \, \nonumber \\
&& \, - \frac{2 v v_4}{\gamma_1} ( \mathbf{p} \times \mathbf{b} )_z \left(
                           \begin{array}{cc}
                             1 & 0 \\
                             0 & -1 \\
                           \end{array}
                         \right) \, \nonumber \\
&& \, + \frac{2 v^2}{\gamma_1^2} ( \mathbf{p} \times \mathbf{b} )_z \left(
                           \begin{array}{cc}
                             U_1 - U_2 - \delta & 0 \\
                             0 & U_1 - U_2 + \delta \\
                           \end{array}
                         \right) \, \nonumber \\
&& \, - \frac{v v_4}{\gamma_1^2} \left(U_1 - U_2\right)\left(
                           \begin{array}{cc}
                           0 & i\pi^{\dagger} \beta^{\dagger} \\
                           -i \pi \beta & 0 \\
                           \end{array}
                           \right)  \nonumber \\
&& \, - \frac{v^2 p^2}{\gamma_1^2} \left(
                           \begin{array}{cc}
                             U_1 - U_2 - \delta & 0 \\
                             0 &  U_2 - U_1 - \delta \\
                           \end{array}
                         \right) \nonumber \\
&& \, + \frac{2 v v_4 p^2}{\gamma_1} \left(
                           \begin{array}{cc}
                             1 & 0 \\
                             0 & 1 \\
                           \end{array}
                         \right)\label{ham2} ,
\end{eqnarray}
where $\beta = b_x + i \xi b_y$, $\beta^{\dagger} = b_x - i \xi b_y$ and $p = |\mathbf{p}|$. Here, we neglect
terms that are quadratic or higher in the magnetic field, cubic or higher in $vp/\gamma_1$ and
cubic or higher in other small parameters $v_4/v$, $\delta / \gamma_1$,
$U_1 / \gamma_1$ and $U_2 / \gamma_1$.

The first term in Eq.~(\ref{ham2}) describes chiral quasiparticles in bilayer graphene \cite{novo06,mccfal06} with the direction of pseudospin in (A1, B2) space lying in the graphene plane and related to that of the electronic momentum,
and this term accounts for a quadratic dispersion $\epsilon \approx v^2 p^2/\gamma_1$.
In the following, we assume the other terms are a small perturbation with respect to this dominant one.
The second term describes different on-site energies $U_1$, $U_2$ on the A1 and B2 sites,
and the third term accounts for trigonal warping due to the presence of skew interlayer coupling
$\gamma_3$ between the A1 and B2 sites. Including parameter $\gamma_3$ doesn't produce magnetic field dependent terms in the Hamiltonian, although it will produce small cross terms in the scattering probability. These will not affect our main results qualitatively so, for simplicity, we will neglect $\gamma_3$.
Instead, magnetic field terms appear due to the inclusion of skew interlayer coupling $\gamma_4$ between A1-A2 or B1-B2 orbitals (fourth term) or due to different on-site energies (fifth term) and there is a cross term, too (sixth term).
The last two terms in Eq.~(\ref{ham2}) are not field dependent but are quadratic in momentum and lead to small corrections to the dispersion.

In the following we neglect terms that are proportional to the unit matrix in (A1, B2) space.
Although they have a small effect on the dispersion relation (parameters $v_4$ and $\delta$ both produce electron-hole asymmetry due to the $p^2$ terms), they do not influence electronic scattering.
In addition, we neglect the small quadratic corrections to the first term in Eq.~(\ref{ham2})
because they do not feature in the results for the scattering rate.
Then, the Hamiltonian may be simplified as
\begin{eqnarray}
H &=& - \frac{v^2}{\gamma_1} \left(
                             \begin{array}{cc}
                               0 & \left(\pi^{\dagger}\right)^2 \\
                               \pi^2 & 0 \\
                             \end{array}
                           \right) \nonumber \\
&& \, + \frac{\Delta}{2} \!\left[ 1 - \frac{2v^2p^2}{\gamma_1^2}\right]\! \left(
       \begin{array}{cc}
         1 & 0 \\
         0 & -1 \\
       \end{array}
     \right)  \, \nonumber \\
&& \, - \frac{2 v^2}{\gamma_1} \!\left[ \frac{v_4}{v} + \frac{\delta}{\gamma_1}\right]\! ( \mathbf{p} \times \mathbf{b} )_z \left(
                           \begin{array}{cc}
                             1 & 0 \\
                             0 & -1 \\
                           \end{array}
                         \right) \, \nonumber \\
&& \, - \frac{v v_4 \Delta}{\gamma_1^2} \left(
                           \begin{array}{cc}
                           0 & i\pi^{\dagger} \beta^{\dagger} \\
                           -i \pi \beta & 0 \\
                           \end{array}
                           \right)  \label{ham3} ,
\end{eqnarray}
where we denote interlayer asymmetry by $\Delta = U_1 - U_2$.
The first magnetic field term takes the form of the Lorentz force and the matrix $\sigma_z$
in the (A1, B2) space causes the pseudospin to acquire a small component perpendicular to the
graphene plane.
The prefactors of this term ($v_4$ and $\delta$) are intrinsic parameters of the lattice and are
thus symmetric with respect to spatial inversion $P [(x, y , z) \rightarrow (-x,-y,-z)]$
and time inversion $T$ so that
$H(\mathbf{p},\mathbf{b}) \xrightarrow{P} H(-\mathbf{p},\mathbf{b})$
and $H(\mathbf{p},\mathbf{b}) \xrightarrow{T} H(-\mathbf{p},-\mathbf{b})$ as expected (note that $\mathbf{b}$ is an axial vector so does not change sign under spatial inversion).
The other magnetic field term (containing $\beta$) is off-diagonal in the (A1, B2) space and it creates a small perturbation of the pseudospin direction within the graphene plane.
Its prefactor contains $\Delta = U_1 - U_2$ which is explicitly odd with respect to spatial inversion so that $H(\mathbf{p},\mathbf{b},\Delta) \xrightarrow{P} H(-\mathbf{p},\mathbf{b},-\Delta)$
and $H(\mathbf{p},\mathbf{b},\Delta) \xrightarrow{T} H(-\mathbf{p},-\mathbf{b},\Delta)$.
In the next Section, we use Hamiltonian~(\ref{ham3}) to determine the correction to the scattering rate caused by the in-plane field.

\subsection{Electron scattering in the presence of an in-plane magnetic field}

Eigenstates $|\mathbf{p}\rangle$ of the two component Hamiltonian~(\ref{ham3})
are used to calculate the scattering rate
$W_{\mathbf{p}^{\prime}\mathbf{p}}$ from state
$|\mathbf{p}\rangle$ to $|\mathbf{p}^{\prime}\rangle$ in the presence of impurities,
keeping up to linear-in-magnetic field terms.
In the presence of a scattering potential $\delta H$, Fermi's golden rule gives
\begin{eqnarray}
W_{\mathbf{p'p}} = \frac{2\pi}{\hbar}\left|\langle \mathbf{p'}\left| \delta H \right| \mathbf{p} \rangle \right|^2 \delta (\epsilon_\mathbf{p}-\epsilon_\mathbf{p'}) \, . \label{formulsp}
\end{eqnarray}
We consider static impurities
\begin{eqnarray*}
\delta H = \sum_{j=1}^{N_i} \hat{Y} u \! \left( \mathbf{r} - \mathbf{R}_j \right) ,
\end{eqnarray*}
where $N_i$ is the number of impurities,
$u \! \left( \mathbf{r} - \mathbf{R}_j \right)$ describes the spatial dependence
of the impurity potential, $\hat{Y}$ is a dimensionless matrix describing structure with respect to the A1, B2 lattice degrees of freedom.
As representative examples, we consider disorder that is symmetric,
$\hat{Y} = \hat{I}$ where $\hat{I}$ is the unit matrix, with equal amounts of scattering
on the two layers, and we consider asymmetric disorder,
$\hat{Y} = (\hat{I} + \zeta\hat{\sigma}_z)/2$, with scattering limited to the lower ($\zeta = 1$)
or upper ($\zeta = -1$) layer.

For symmetric disorder $\hat{Y} = \hat{I}$, we find
\begin{eqnarray}
\delta W_{\mathbf{p}^{\prime}\mathbf{p}}^{(\mathrm{s})} &=& \frac{2\pi}{\hbar}
\frac{n_i}{L^2} \left| \tilde{u} \! \left( \mathbf{p}^{\prime} - \mathbf{p} \right) \right|^2
\delta \! \left( \epsilon_{\mathbf{p}} - \epsilon_{\mathbf{p}^{\prime}} \right) \label{scatts} \\
&& \times \bigg\{
\tfrac{1}{2} \left( 1 + \cos [2(\phi^{\prime} - \phi)] \right)  \nonumber \\
&& \quad + \frac{v_4 \Delta}{2 \gamma_1 vp^2} \sin [2(\phi^{\prime} - \phi)]
( \mathbf{p}^{\prime} - \mathbf{p} ) \cdot \mathbf{b} \nonumber \\
&& \quad - \bigg[\frac{\Delta \gamma_1}{2 v^2 p^4}
\!\!\left( \frac{v_4}{v} + \frac{\delta}{\gamma_1} \right)\!\!
\left( 1 - \frac{2v^2p^2}{\gamma_1^2} \right) \nonumber \\
&&\quad\times \left( 1 - \cos [2(\phi^{\prime} - \phi)] \right)
\![( \mathbf{p}^{\prime} + \mathbf{p} ) \times \mathbf{b}]_z \bigg]\bigg\}  . \nonumber
\end{eqnarray}
where the density of impurities is $n_i = N_i/L^2$,
and we keep terms up to linear in magnetic field and linear in $\Delta = U_1 - U_2$.
We neglect interference between different impurities and use the Fourier transform of the impurity potential
\begin{eqnarray*}
\tilde{u} \! \left( \mathbf{q} \right) = \int d^2r \, u \! \left( \mathbf{r} \right)
e^{-i\mathbf{q}.\mathbf{r}/\hbar}
\, .
\end{eqnarray*}
For asymmetric disorder $\hat{Y} = (\hat{I} + \zeta\hat{\sigma}_z)/2$, we find
\begin{eqnarray}
\delta W_{\mathbf{p}^{\prime}\mathbf{p}}^{(\mathrm{a})} &=& \frac{2\pi}{\hbar}
\frac{n_i}{L^2} \left| \tilde{u} \! \left( \mathbf{p}^{\prime} - \mathbf{p} \right) \right|^2
\delta \! \left( \epsilon_{\mathbf{p}} - \epsilon_{\mathbf{p}^{\prime}} \right) \label{scatta} \\
&& \!\!\!\!\!\! \!\!\!\!\!\! \times \bigg\{
\frac{1}{4}
+ \frac{s \zeta \Delta \gamma_1}{4 v^2 p^2}\!\left(1 - \frac{2v^2p^2}{\gamma_1^2}\right) \nonumber \\
&& \!\!\!\!\!\! \!\!\!\!\!\!
- \left( \frac{v_4}{v} + \frac{\delta}{\gamma_1} \right)\!\! \!\left( s \zeta - \frac{\Delta}{\gamma_1} + \frac{\Delta \gamma_1}{2v^2p^2} \right) \!\!
\frac{[( \mathbf{p}^{\prime} + \mathbf{p} ) \times \mathbf{b} ]_z}{2  p^2} \bigg\} , \nonumber
\end{eqnarray}
where $s=+1$ ($s=-1$) for states in the conduction (valence) band.

The linear-in-field parts of the scattering rates Eqs.~(\ref{scatts},\ref{scatta}),
which are relevant for the magnetic ratchet effect,
may be written in a general form as
\begin{eqnarray}
&& \delta W_{\mathbf{p}^{\prime}\mathbf{p}} = \frac{1}{L^2} \left| \tilde{u} \! \left( \mathbf{p}^{\prime} - \mathbf{p} \right) \right|^2
\delta \! \left( \epsilon_{\mathbf{p}} - \epsilon_{\mathbf{p}^{\prime}} \right) \label{wgen} \\
&& \!\!\!\!\times \bigg\{ \!\!
\left( \Omega - \Omega_c \cos [2(\phi^{\prime} - \phi)] \right) \!\!
\left[ B_x \left( p_y^{\prime} + p_y \right) - B_y \left( p_x^{\prime} + p_x \right) \right] \nonumber \\
&&
\quad  + \, \Omega_s \sin [2(\phi^{\prime} - \phi)] \!\!
\left[ B_x \left( p_x^{\prime} - p_x \right) + B_y \left( p_y^{\prime} - p_y \right) \right] \!\! \bigg\} \, , \nonumber
\end{eqnarray}
where the angle-independent factors $\Omega$, $\Omega_c$, $\Omega_s$ are:
\begin{eqnarray*}
\Omega_c^{(\mathrm{s})} = \Omega^{(\mathrm{s})}
&=& \frac{\pi e d n_i \Delta \gamma_1}{2\hbar v^2 p^4} \!\!
\left( \frac{\gamma_4}{\gamma_0} + \frac{\delta}{\gamma_1} \right) \!\!
\left( 1 - \frac{2v^2p^2}{\gamma_1^2} \right) \, , \\
\Omega_s^{(\mathrm{s})} &=& \frac{\pi e d n_i \Delta \gamma_4}{2\hbar p^2 \gamma_1 \gamma_0} \, ,
\end{eqnarray*}
for symmetric disorder, and for asymmetric disorder:
\begin{eqnarray*}
\Omega^{(\mathrm{a})}
&=& \frac{\pi e d n_i }{2\hbar p^2} \!\!
\left( \frac{\gamma_4}{\gamma_0} + \frac{\delta}{\gamma_1} \right) \!\!
\left( s \zeta - \frac{\Delta}{\gamma_1} + \frac{\Delta \gamma_1}{2v^2p^2} \right) \, , \\
\Omega_c^{(\mathrm{a})} = \Omega_s^{(\mathrm{a})} &=& 0 \, .
\end{eqnarray*}

For symmetric disorder, symmetry must be broken by interlayer asymmetry $\Delta = U_1 - U_2$ in order to produce
magnetic-field-dependent terms in the scattering rate.
The term proportional to $( \mathbf{p}^{\prime} - \mathbf{p} ) \cdot \mathbf{b}$ in the
scattering rate, Eq.~(\ref{scatts}), arises directly from the off-diagonal term in the Hamiltonian
[the final term in Eq.~(\ref{ham3})], whereas the term proportional to
$[( \mathbf{p}^{\prime} + \mathbf{p} ) \times \mathbf{b}]_z$ arises from the interplay of the Lorentz-force-term
and the field-independent $\Delta$ term [the second and third terms in Eq.~(\ref{ham3})].
By contrast, for asymmetric disorder, interlayer asymmetry is not essential and, in fact,
the leading term in the scattering rate, Eq.~(\ref{scatta}), arises from the Lorentz-force-term in the Hamiltonian which does not contain $\Delta$ [the third term in Eq.~(\ref{ham3})].

\subsection{The Boltzmann equation for the rachet current in a two-dimensional material}

Ratchet current can be induced by an in-plane magnetic field in any two-dimensional electron
system with $z \rightarrow - z$ asymmetry and an arbitrary isotropic dispersion $\epsilon_p$.
We use a degeneracy factor $g$ ($g=4$ for spin and
valley in graphene), group velocity
\begin{eqnarray*}
v_g = \frac{d\epsilon}{dp} \, ,
\end{eqnarray*}
and density of states per spin and per valley, per unit area, $\Gamma = p/2 \pi \hbar^2 v_g$.
For a spatially homogeneous sample, we consider the Boltzmann equation
\begin{eqnarray}
- e\mathbf{E}_{\parallel}.\nabla_p f (\mathbf{p},t)
+ \frac{\partial f (\mathbf{p},t)}{\partial t}
= S\!\!\:\{f\} , \label{boltz1}
\end{eqnarray}
where the electron distribution $f (\mathbf{p},t)$ is a function of
momentum $\mathbf{p}$ and time $t$,
the in-plane ac field is
$\mathbf{E}_{\parallel}(t) = \mathbf{E}_{\parallel} e^{-i\omega t} + \mathbf{E}_{\parallel}^{\ast} e^{i\omega t}$,
and the electronic charge is $-e$, $e>0$.
The collision integral $S\!\!\:\{f\}$ is given by
\begin{eqnarray}
S\!\!\:\{f\} = \sum_{\mathbf{p}^{\prime}} \left[ W_{\mathbf{p}\mathbf{p}^{\prime}}
f (\mathbf{p}^{\prime},t) - W_{\mathbf{p}^{\prime}\mathbf{p}}
f (\mathbf{p},t) \right] , \label{colint}
\end{eqnarray}
where $W_{\mathbf{p}^{\prime}\mathbf{p}}$ is the scattering rate~Eq.~(\ref{wgen}).
In the scattering rate, we perform an harmonic expansion of the impurity potential,
\begin{eqnarray*}
\left| \tilde{u} \! \left( \mathbf{p}^{\prime} - \mathbf{p} \right) \right|^2
= \sum_m \nu_m e^{im (\phi^{\prime} - \phi)} \, ,
\end{eqnarray*}
where $\phi$ is the polar angle of momentum. The expansion has
the constraint that $\nu_{-m} = \nu_m$ as it is an even function of
$(\phi^{\prime} - \phi)$.

The Boltzmann equation Eq.~(\ref{boltz1}) is written in terms of polar coordinates $(p,\phi)$ for momentum and the distribution function is expanded in terms of $\phi$ and $t$ harmonics with coefficients $f_m^{(n)}$:
\begin{eqnarray}
f (\mathbf{p},t) = \sum_{n,m} f_m^{(n)} e^{im\phi - i n \omega t} ,
\end{eqnarray}
where $m$, $n$ are integers.
Multiplying the Boltzmann equation by a factor $\exp ( - i j \phi + i \ell \omega t )$,
where $j$, $\ell$ are integers, and integrating over a period $2\pi$ of angle $\phi$
and a period of time $t$, leads to coupled equations between different harmonic coefficients:
\begin{eqnarray}
f_j^{(\ell)} \! \left( \tau_{|j|}^{-1} - i \ell \omega \right)
&=& \alpha_{j-1} f_{j-1}^{(\ell - 1)} + \eta_{j+1} f_{j+1}^{(\ell - 1)} \label{coupled} \\
&& \!\!\!
+ \tilde{\alpha}_{j-1} f_{j-1}^{(\ell + 1)} + \tilde{\eta}_{j+1} f_{j+1}^{(\ell + 1)} + \delta \!S_{j}^{(\ell)}  , \nonumber
\end{eqnarray}
where the scattering rate (in the absence of magnetic field) is defined as
\begin{eqnarray}
\tau_{|j|}^{-1} &=& \frac{2\pi}{\hbar} \sum_{\mathbf{p}^{\prime}}
\left| \langle \mathbf{p}^{\prime} | \delta H | \mathbf{p} \rangle \right|^2
\delta \! \left( \epsilon_{\mathbf{p}} - \epsilon_{\mathbf{p}^{\prime}} \right) \nonumber \\
&& \qquad \qquad \qquad \times \left[ 1 - \cos \left( j \left[ \phi^{\prime} - \phi \right] \right) \right] \, . \label{tau}
\end{eqnarray}
In the following we write $\tau_1 = \tau$ for simplicity.
Operators arising from the electric field in Eq.~(\ref{boltz1}) are
\begin{eqnarray*}
\alpha_{j} &=& \frac{e \left(E_x - i E_y \right)}{2} \left( - \frac{j}{p} + \frac{\partial}{\partial p} \right) \, , \\
\tilde{\alpha}_{j} &=& \frac{e \left(E_x^{\ast} - i E_y^{\ast} \right)}{2} \left( - \frac{j}{p} + \frac{\partial}{\partial p} \right) \, , \\
\eta_{j} &=& \frac{e \left(E_x + i E_y \right)}{2} \left( \frac{j}{p} + \frac{\partial}{\partial p} \right) \, , \\
\tilde{\eta}_{j} &=& \frac{e \left(E_x^{\ast} + i E_y^{\ast} \right)}{2} \left( \frac{j}{p} + \frac{\partial}{\partial p} \right) \, . \\
\end{eqnarray*}
Factors $\delta \!S_{j}^{(\ell)}$ in Eq.~(\ref{coupled}) describe the correction to scattering caused by the magnetic field, the relevant ones have small values of $j$:
\begin{eqnarray*}
\delta \!S_{0}^{(\ell)} &=& 0 , \\
\delta \!S_{1}^{(\ell)} &=& \tfrac{1}{2} p \, \Gamma (\epsilon) \left( B_y - i B_x \right)
\Lambda_1 f_{2}^{(\ell)} , \\
\delta \!S_{-1}^{(\ell)} &=& \tfrac{1}{2} p \, \Gamma (\epsilon) \left( B_y + i B_x \right)
\Lambda_1 f_{-2}^{(\ell)} \, \\
\delta \!S_{2}^{(\ell)} &=& \tfrac{1}{2} p \, \Gamma (\epsilon) \left[ \left( B_y + i B_x \right) \Lambda_1 f_{1}^{(\ell)}
+ \left( B_y - i B_x \right) \Lambda_2 f_{3}^{(\ell)} \right] , \\
\delta \!S_{-2}^{(\ell)} &=& \tfrac{1}{2} p \, \Gamma (\epsilon) \left[ \left( B_y - i B_x \right) \Lambda_1 f_{-1}^{(\ell)}
+ \left( B_y + i B_x \right) \Lambda_2 f_{-3}^{(\ell)} \right] ,
\end{eqnarray*}
where
\begin{eqnarray}
\Lambda_1 \!\! &=& \!\! \Omega (\nu_0 - \nu_2) + \tfrac{1}{2} \Omega_c (\nu_0 - 2\nu_2 + \nu_4) \nonumber \\
&& \qquad + \tfrac{1}{2} \Omega_s (\nu_0 - 2\nu_1 + 2\nu_3 - \nu_4) \, , \label{lam1} \\
\Lambda_2 \!\! &=& \!\! \Omega (\nu_0 + \nu_1 - \nu_2 - \nu_3)
+ \tfrac{1}{2} \Omega_c (\nu_0 - 2\nu_2 - \nu_3 + \nu_4 + \nu_5) \nonumber \\
&& \qquad - \tfrac{1}{2} \Omega_s (\nu_0 - \nu_3 - \nu_4 + \nu_5) \, . \label{lam2}
\end{eqnarray}

\begin{figure}[ht]
   \centering
   \includegraphics[scale=0.25]{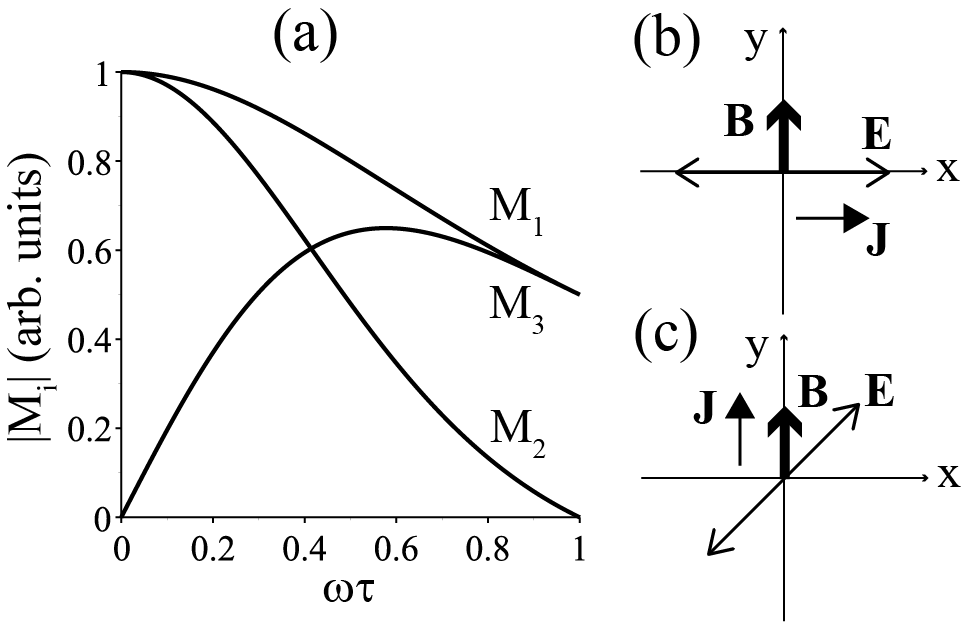}
   \caption {(a) frequency dependence of the coefficients $M_i$, Eq.~(\ref{mcoeffs}),
   characterizing the polarization dependence of the ratchet current Eq.~(\ref{jden}).
   For simplicity, we set $\tau_2 = \tau$. (b) and (c) illustrate the direction of the
   ratchet current $\mathbf{J}$ in the graphene ($x$-$y$) plane for linear polarization (and $M_1>0$) in (b) the $x$ direction, and (c) at $45^{\circ}$ to the $x$-axis.}
    \label{fig:pol}
\end{figure}

The dc current density is given by
\begin{eqnarray*}
\mathbf{J} = - \frac{g}{L^2} \sum_{\mathbf{p}} e \mathbf{v}_g \delta \!f ,
\end{eqnarray*}
where $\delta \!f = f_1^{(0)} e^{i\phi} + f_{-1}^{(0)} e^{-i\phi}$
is the only part of the time-independent distribution that gives a non-zero contribution
after integrating over all angles $\phi$.
The coupled equations~(\ref{coupled}) for the harmonics $f_j^{(\ell)}$
of the distribution function are used to express $\delta \!f$ in terms of
the equilibrium distribution $f_0^{(0)}$ and, assuming a degenerate electron gas,
we find the current density has the form~(\ref{jden}) with
\begin{eqnarray}
M_1 \!\!&=& \!\! - \frac{g e^3}{16 \pi^2 \hbar^4} \frac{1}{1 + \omega^2 \tau^2} \!
\left( 4 \Lambda_1 p^2 \tau^2 \tau_2 + v_g p^3 \tau ( \Lambda_1 \tau \tau_2 )^{\prime}\right) \! , \nonumber \\
M_2 \!\!&=& \!\! \frac{g e^3}{16 \pi^2 \hbar^4}
\frac{\Lambda_1 p^2 \tau^2\tau_2 \left( 1 - \omega^2 \tau \tau_2 \right)}
{\left( 1 + \omega^2 \tau^2 \right) \left( 1 + \omega^2 \tau_2^2 \right)}
\! \left( 1 - p v_g^{\prime} - \frac{v_g p \tau^{\prime}}{\tau} \right) \! , \nonumber \\
M_3 \!\!&=& \!\! - \frac{g e^3}{16 \pi^2 \hbar^4}
\frac{\Lambda_1 p^2 \tau^2 \tau_2 \omega \left( \tau + \tau_2 \right)}
{\left( 1 + \omega^2 \tau^2 \right) \left( 1 + \omega^2 \tau_2^2 \right)}
\! \left( 1 - p v_g^{\prime} - \frac{v_g p \tau^{\prime}}{\tau} \right) \! , \nonumber \\
\label{mcoeffs}
\end{eqnarray}
where $(\ldots)^{\prime} \equiv \partial (\ldots) / \partial \epsilon$ and
all parameters are evaluated on the Fermi surface.
These equations, which generalize those in Refs.~\cite{tara11,drex13}, describe the ratchet effect in a two-dimensional material with isotropic dispersion. Parameters such as the scattering times $\tau$, $\tau_2$, Eq.~(\ref{tau}), and $\Lambda_1$, Eq.~(\ref{lam1}), describing the effect of in-plane magnetic field will be specific to the given material, in this case bilayer graphene.
The frequency dependence of the $M_i$ coefficients is plotted in Fig.~\ref{fig:pol}(a)
for the case $\tau_2 = \tau$.

\section{Ratchet effect in bilayer graphene}

In the following, we assume that the dispersion of bilayer graphene is quadratic
$\epsilon = v^2p^2/\gamma_1$ so that $v_g = 2v^2p/\gamma_1$ (note that $v_g \neq v$ where $v$ is the group velocity of monolayer graphene). In this case, the factor
$1 - p v_g^{\prime} - v_g p \tau^{\prime}/\tau$ simplifies to
$- v_g p \tau^{\prime}/\tau$, with the result that the existence of non-zero $M_2$ and $M_3$ relies on the energy
dependence of the scattering rate $\tau$, irrespective of the effect of the in-plane field.

In bilayer graphene, overscreened Coulomb impurities act like short-range scatterers \cite{adamdassarma08}:
\begin{eqnarray*}
u \! \left( \mathbf{r} - \mathbf{R}_j \right) = u_0 \delta \! \left( \mathbf{r} - \mathbf{R}_j \right) ,
\end{eqnarray*}
in which case $\tilde{u} \! \left( \mathbf{p}^{\prime} - \mathbf{p} \right) = u_0$
and the scattering rate Eq.~(\ref{tau}) simplifies as
\begin{eqnarray*}
\mathrm{symmetric\,\,disorder\!:} \quad \tau^{-1} &=& 2 \tau_2^{-1} = \frac{n_i u_0^2 \gamma_1}{4 \hbar^3 v^2} \, , \\
\mathrm{asymmetric\,\,disorder\!:} \quad \tau^{-1} &=& \tau_2^{-1} = \frac{n_i u_0^2 \gamma_1}{8 \hbar^3 v^2} \, . \\
\end{eqnarray*}
Furthermore, if $u_0$ is independent of energy, then so is $\tau$ and $M_2 = M_3 = 0$.
As the potential is isotropic, $\nu_0 = u_0^2$ is the only non-zero harmonic and
parameter $\Lambda_1$ simplifies as
$\Lambda_1 = u_0^2 [ \Omega + (\Omega_c + \Omega_s)/2]$.
For symmetric disorder, Eq.~(\ref{scatts}), we find
\begin{eqnarray}
M_1^{(\mathrm{s})} = \frac{e^4 d \tau^2}{2\pi \hbar^2 m (1 + \omega^2 \tau^2)} \frac{\Delta}{\gamma_1}
\left( 5\frac{\gamma_4}{\gamma_0} + 6\frac{\delta}{\gamma_1} \right)  , \label{ms}
\end{eqnarray}
where the mass $m = \gamma_1 / 2v^2$.
For asymmetric disorder, Eq.~(\ref{scatta}), we find
\begin{eqnarray}
M_1^{(\mathrm{a})} = - \frac{e^4 d \tau^2}{\pi \hbar^2 m (1 + \omega^2 \tau^2)} \!\!
\left( \frac{\gamma_4}{\gamma_0} + \frac{\delta}{\gamma_1} \right)\!\!\!\left( s \zeta - \frac{\Delta}{\gamma_1} \right) \! . \label{ma}
\end{eqnarray}
These expressions are independent of Fermi level other than through the factor $s=\pm 1$ for conduction/valence bands.
Linear dependence on gate-induced interlayer asymmetry $\Delta / \gamma_1$ occurs for symmetric disorder because interlayer asymmetry is required to break symmetry in this case, whereas interlayer asymmetry $\Delta / \gamma_1$ is a small correction for asymmetric disorder (as $\Delta / \gamma_1 \ll 1$ in general).
Note that for linear polarization $E_x = E_0 \cos \theta$, $E_y = E_0 \sin \theta$ where $E_0$ is real, and $M_2 = M_3 = 0$,
then the expression for the ratchet current~(\ref{jden}) simplifies as
\begin{eqnarray*}
J_x &=& M_1 E_0^2 \left( B_y \cos 2\theta - B_x \sin 2 \theta \right) \, , \\
J_y &=& M_1 E_0^2 \left( B_x \cos 2\theta + B_y \sin 2 \theta \right) \, .
\end{eqnarray*}
This indicates that the direction of the ratchet current is given by a rotation of the magnetic field
direction by an amount determined by the polarization angle $\theta$, as illustrated in
Fig.~\ref{fig:pol}(b) and (c).

To estimate the magnitude of the ratchet current, we use
parameter values determined by infrared spectroscopy~\cite{zhang08} (similar values were measured by Ref.~\cite{kuz09}, too) which include
$\gamma_0 = 3.0\,$eV, $\gamma_1 = 0.4\,$eV, $\gamma_4 = 0.015\,$eV, $\delta = 0.018\,$eV, and we also use interlayer spacing $d \approx 3.3\,${\AA} and mass $m \approx 0.05m_e$ where $m_e$ is the free electron mass.
We compare the magnitude of $M_1^{(\mathrm{a})}$ for asymmetric disorder Eq.~(\ref{ma}) with the theoretical prediction of Ref.~\cite{drex13} for $\pi$-$\sigma$ orbital hybridization in hydrogenated monolayer graphene
$M_1^{(\mathrm{mon})} = 12 e^4 z_{\pi\sigma} \epsilon_F \tau^2 / [\pi \hbar^2 m_e \epsilon_{\pi\sigma}(1 + \omega^2 \tau^2)]$
(Eq.(7) in Ref.~\cite{drex13})
where $z_{\pi\sigma} \approx 0.15\,${\AA} is the distance between $\pi$-$\sigma$ orbitals,
$\epsilon_{\pi\sigma} \approx 10\,$eV is the energy between $\pi$-$\sigma$ orbitals,
$\epsilon_F \approx 150\,$meV is the Fermi level.
Assuming the scattering times $\tau$ in bilayer and monolayer are the same order of magnitude,
we find
$|M_1^{(\mathrm{a})} / M_1^{(\mathrm{mon})}| \sim
(d / z_{\pi\sigma} ) (m_e / m) ( \gamma_4/\gamma_0 + \delta/\gamma_1 )
/(12 \epsilon_F / \epsilon_{\pi\sigma})
\sim 20 \times 20 / 4 \sim 100$, indicating that the magnitude of the ratchet effect
in bilayer graphene should be substantial.

\section{Conclusions}

We have considered the orbital effect of an in-plane magnetic field on electrons in
bilayer graphene.
Previously the orbital effect of an in-plane magnetic field on the electronic spectrum was modeled
\cite{persh10,roy13,don16} using the so-called minimal tight-binding model which includes only intralayer and vertical interlayer coupling, accounting for quadratic in magnetic field terms in the low-energy Hamiltonian. At low energy, these terms have a similar effect as homogeneous lateral strain in producing a change in topology of the band structure \cite{mucha11,son11,mar12,he14}, although, owing to the small interlayer distance, a huge magnetic field of magnitude~$\sim 100\,$T would be required to observe this.
Here, we derived linear-in-field terms in the Hamiltonian Eq.~(\ref{ham3})
arising from skew interlayer coupling and non-uniform on-site energies.
We found two types of term, the first has the form of the Lorentz force and it causes the
pseudospin (the relative amplitude of the wave function on the two layers) to
acquire a small component perpendicular to the graphene plane,
the second term is off-diagonal in the layer-space and it creates a small perturbation of the pseudospin direction within the graphene plane.
We modeled the influence of these terms on electronic scattering and their manifestation in the magnetic ratchet effect.
We estimate that the effect should be substantial, two orders of magnitude greater
than that in monolayer graphene~\cite{drex13},
as well as being sensitive to the nature of disorder and tunable by gate voltage.


\begin{acknowledgments}
We thank X.~Chen and J.~R.~Wallbank for useful discussions.
This work was funded by the EU Flagship Project and the ERC Synergy Grant Hetero2D.
\end{acknowledgments}

\end{document}